\documentclass[reprint,pra,twocolumn,showpacs,superscriptaddress,aps,longbibliography]{revtex4-2}

\usepackage{amsmath,amssymb,amsfonts}
\usepackage{graphicx}
\usepackage{braket}
\usepackage{multirow}
\usepackage{float}
\usepackage{rotating}

\usepackage{adjustbox}
\usepackage{xcolor}
\usepackage[dvipsnames]{xcolor}
\usepackage{mathdots}
\usepackage{microtype}

\usepackage[colorlinks]{hyperref}
\hypersetup{%
        plainpages=true,
        breaklinks=true,
        hypertexnames=false,
        pageanchor=true,
        colorlinks=true,
        linkcolor={blue},
        citecolor={blue},
        urlcolor={blue},
        anchorcolor={black}
      }

\begin{document}

%=========================================================
% Title
%=========================================================

\title{Structured Quantum State Reconstruction via Physically Motivated Operator Selection}

\author{Ayush Chambyal}
\affiliation{Department of Physics and Astronomical Science, Central University of Himachal Pradesh, Kangra, H.P. 176206, India}

\author{Brijesh}
\affiliation{Department of Physics and Astronomical Science, Central University of Himachal Pradesh, Kangra, H.P. 176206, India}

\author{Rakesh Sharma}
\email[Corresponding author: ]{rakeshiisermohali@hpcu.ac.in}
\affiliation{Department of Physics and Astronomical Science, Central University of Himachal Pradesh, Kangra, H.P. 176206, India}
\date{\today}

%=========================================================
% Abstract
%=========================================================

\begin{abstract}
Quantum state tomography (QST) scales exponentially in both measurement and computational cost, making full reconstruction impractical for multi-qubit systems. Existing approaches attempt to reduce this complexity, but do not explicitly restrict the operator space based on physically relevant correlations.
We develop a structured QST framework in which the density matrix is reconstructed using a restricted set of observables in a Gibbs representation. The Structured Gibbs Quantum State Tomography (SG-QST) is built by progressively including local, nearest-neighbor, and global correlations. Benchmarking on three, four, and five-qubit. GHZ states shows that comparable fidelity can be achieved with significantly fewer parameters by restricting the operator space to physically relevant observables.
These results demonstrate that physically motivated operator-space restriction enables efficient and interpretable quantum state reconstruction.

\end{abstract}

\maketitle

%=======================================================
%=========================================================
\section{Introduction}

The characterization of quantum states is a central task in quantum information science, playing a key role in the verification and validation of quantum devices and protocols~\cite{nielsen2002quantum,eisert2020quantum}. In experimental settings, the quantum state is not directly accessible and must be inferred from measurement data through quantum state tomography (QST), where the density matrix $\rho$ is reconstructed from expectation values of measured observables~\cite{paris2004quantum,lvovsky2009continuous, Gaikwad-qst-2025}.

A fundamental limitation of QST is the exponential scaling of the parameter space with system size. A general $n$-qubit quantum state is described by $4^n - 1$ real parameters~\cite{nielsen2002quantum,haah2017sample, patel-prr-2026}, making full state reconstruction increasingly demanding in both measurement and computational resources as the number of qubits grows. Standard reconstruction techniques such as linear inversion~\cite{james2001measurement} provide direct estimates but may yield non-physical density matrices, while maximum likelihood estimation (MLE)~\cite{hradil1997quantum,rehacek2007diluted, Gaikwad-qip-2021} enforces positivity and trace constraints without addressing the exponential scaling of the problem. 

To mitigate these challenges, several approaches exploit structural properties of quantum states to reduce measurement or computational complexity. Compressed sensing methods enable reconstruction of approximately low-rank states from a reduced number of measurements under suitable conditions~\cite{Gaikwad-qip-2022, gross2010quantum,flammia2012quantum,kalev2015quantum}. Tensor-network-based approaches allow efficient tomography of many-body systems that obey area-law entanglement scaling~\cite{cramer2010efficient,lanyon2017efficient}. Bayesian methods incorporate prior information and adaptive measurement strategies to improve estimation efficiency~\cite{huszar2012adaptive}, while neural-network-based approaches provide flexible variational representations for quantum states~\cite{gaikwad-pra-2024, torlai2018neural,carleo2017solving}.

Despite these advances, each approach relies on specific assumptions about the underlying quantum state. Compressed sensing assumes approximate low-rank structure~\cite{flammia2012quantum}, tensor-network methods are efficient primarily for states with restricted entanglement growth~\cite{cramer2010efficient}, and Bayesian approaches depend on the availability of suitable prior information~\cite{huszar2012adaptive}. Neural-network representations, while expressive, often lack direct interpretability in terms of physically transparent observables.
An alternative direction focuses on estimating properties of quantum states without reconstructing the full density matrix. Direct fidelity estimation enables efficient evaluation of overlaps with a target state using a subset of Pauli measurements~\cite{flammia2011direct,da2011practical}, while classical shadow tomography allows prediction of many observables from randomized measurements with provable guarantees~\cite{huang2020predicting,aaronson2018shadow,huang2021efficient}. However, these approaches do not provide an explicit and compact representation of the reconstructed state.

These limitations motivate approaches that incorporate physically relevant structure directly into the reconstruction procedure. In many quantum systems, correlations are not uniformly distributed over the full operator space but are instead concentrated in specific subsets of observables. This idea is closely related to quantum state reconstruction from incomplete data and maximum entropy principles, where the density matrix is inferred from a restricted set of constraints~\cite{teo2012incomplete,gaikwad-epjd-2023}.

A paradigmatic example is provided by Greenberger--Horne--Zeilinger (GHZ) states~\cite{greenberger1989going}, which exhibit strong multipartite entanglement and nonlocal correlations~\cite{dur2000three,guhne2009entanglement}. For such states, single-qubit reduced density matrices are maximally mixed, while the essential information is encoded in global multi-qubit correlations, indicating that the relevant structure of the state can be captured by a restricted set of observables. This perspective is consistent with the principle of maximum entropy~\cite{jaynes1957information}, which constructs the least-biased quantum state compatible with known expectation values. In the quantum setting, this leads to a Gibbs-form density matrix that ensures physicality while allowing controlled selection of the operator set~\cite{petz2008quantum}.

In this work, we develop a Structured Gibbs Quantum State Tomography (SG-QST), framework in which the operator space is explicitly restricted based on physically motivated observables. Unlike standard maximum entropy approaches that rely on arbitrary constraint selection, we construct a systematic hierarchy of Gibbs models that progressively incorporate local, short-range, and global correlations.

We benchmark this approach against standard full reconstruction methods, including MLE and positive semidefinite (PSD) projection, for three, four, and five-qubit GHZ states.G HZ states represent a natural and well-motivated choice for this purpose, as they are among the most extensively studied multipartite entangled states~\cite{greenberger1989going, dur2000three, guhne2009entanglement}, with established importance in quantum communication, quantum cryptography, and foundational tests of quantum nonlocality~\cite{pan2012, hillery1999}. Their experimental realization across photonic systems, trapped ions, and superconducting qubits has made them a standard benchmark for quantum protocols and devices~\cite{pan2000, monz2011, song2017}. Furthermore, their highly structured correlation pattern, in which 
all physically relevant information is encoded entirely in global multi-qubit observables rather than local terms~\cite{dur2000three, guhne2009entanglement}, provides a clear and controlled setting for evaluating the effectiveness of physically motivated operator-space restriction~\cite{guhne2009entanglement}.  

Our results demonstrate that by targeting these dominant correlations, high-fidelity reconstruction can be achieved using a substantially reduced number of parameters, with SG-QST complexity determined by the correlation structure of the state rather than the dimension of the Hilbert space. These findings establish SG- QST as a scalable and physically interpretable framework for quantum state reconstruction.

We present the theoretical background including the GHZ state formalism and the Gibbs (exponential) parametrization of the density matrix~\cite{jaynes1957information,petz2008quantum}. Then reconstruction methodology was described in detail, covering the full reference methods --- linear inversion with positive semidefinite projection~\cite{james2001measurement} and maximum likelihood estimation~\cite{hradil1997quantum,rehacek2007diluted}, as well as the SG--QST. The 
results for three, four, and five-qubit GHZ states~\cite{greenberger1989going,pan2000,monz2011}, including fidelity analysis, agreement with MLE, observable reconstruction error, residual analysis, and scaling behavior across system sizes were shown. Finally,the main conclusions and discusses future directions for extending structured quantum state tomography to larger and more complex quantum systems.

%=========================================================
\section{Theoretical Framework}
\label{sec:Theoretical Framework}
The state of an $n$-qubit quantum system is described by a density matrix $\rho \in \mathbb{C}^{2^n \times 2^n}$, which is Hermitian, positive semidefinite, and normalized such that $\mathrm{Tr}(\rho) = 1$~\cite{nielsen2002quantum}. Observable quantities are obtained through expectation values $\langle O \rangle = \mathrm{Tr}(\rho O)$, which form the basis of experimental quantum state characterization.

In this work, we focus on Greenberger–Horne–Zeilinger (GHZ) states~\cite{greenberger1989going}, defined as
\begin{equation}
\ket{\mathrm{GHZ}_n} = \frac{1}{\sqrt{2}} \left( \ket{0}^{\otimes n} + \ket{1}^{\otimes n} \right).
\end{equation}
The corresponding density matrix is given by
\begin{equation}
\rho_{\mathrm{GHZ}} =
\frac{1}{2} \left(
\ket{0^n}\bra{0^n}
+ \ket{1^n}\bra{1^n}
+ \ket{0^n}\bra{1^n}
+ \ket{1^n}\bra{0^n}
\right)
\end{equation}
which contains diagonal population terms and off-diagonal coherence terms representing global quantum correlations.
When expressed in the Pauli operator basis, any $n$-qubit state can be written as
\begin{equation}
\rho = \frac{1}{2^n} \sum_{P \in \mathcal{P}_n} \langle P \rangle P,
\end{equation}
where $\mathcal{P}_n = \{I, X, Y, Z\}^{\otimes n}$ is the set of Pauli strings and $\langle P \rangle = \mathrm{Tr}(\rho P~$\cite{nielsen2002quantum}. GHZ states exhibit a distinctive structure in this representation: local expectation values vanish, while global correlations such as
\begin{equation}
\langle X^{\otimes n} \rangle = 1
\end{equation}
capture the essential coherence of the state~\cite{guhne2009entanglement}. This indicates that the dominant physical information is encoded in higher-order correlations rather than local observables.

We therefore parameterize the density matrix using a Gibbs (exponential)form~\cite{jaynes1957information, petz2008quantum}
\begin{equation}
\rho = \frac{e^{-H}}{\mathrm{Tr}(e^{-H})}, \
\end{equation}
This parametrization ensures positivity and unit trace by construction. The effective Hamiltonian is written as 
\begin{equation}
\quad H = \sum_k \lambda_k P_k
\end{equation}
where $\{P_k\}$ is a selected set of Pauli operators chosen to reflect the known correlation structure of GHZ states, where local expectation values vanish and global multi-qubit correlation dominate. As a result,only these observables contribute to the reconstruction. Since the Hamiltonian is constructed from this restricted operator set, the resulting density matrix is constrained to the corresponding operator subspace, directly controlling the structure and complexity of the reconstruction.

These observations motivate a structured approach to quantum state reconstruction, in which the operator space is systematically restricted based on the underlying physical correlations of the state. Rather than treating all observables equally, this approach prioritizes those that capture the dominant nonlocal features, particularly the global coherence characteristic of GHZ states. In the following section, we formalize this approach and introduce the corresponding hierarchy of reconstruction models.
\section{Methodology}
\label{sec:Methodology}
For structured states such as GHZ states, the physically relevant information is not uniformly distributed across the operator space, but is instead concentrated in specific correlation sectors, most notably global multi-qubit coherence~\cite{greenberger1989going, guhne2009entanglement}. 

This indicates that accurate reconstruction can be achieved without accessing the full operator space, provided that the selected observables capture the dominant physical correlations of the state.

\subsection{Full Reconstruction Methods}

As reference approaches, we employ two standard reconstruction techniques: linear inversion followed by positive semi-definite (PSD) projection, and maximum likelihood estimation (MLE).

In linear inversion, expectation values of Pauli operators are used to reconstruct the density matrix in the Pauli basis~\cite{james2001measurement}. Since this procedure does not guarantee a physical state, the resulting matrix is projected onto the space of positive semi-definite density matrices with unit trace.

In contrast, MLE enforces physicality by construction. The density matrix is parameterized as~\cite{hradil1997quantum}, 
\begin{equation}
\rho =  \frac{T^\dagger T}{ \mathrm{Tr(T^\dagger T)}}
\end{equation}
 and the parameters are obtained by minimizing the deviation between predicted and measured expectation values. Due to the non-convex nature of this optimization, multiple random initializations are employed, and the solution with the lowest loss is selected as the maximum likelihood estimate. The optimization is performed using the L-BFGS-B quasi-Newton algorithm~\cite{byrd1995limited}.

Both PSD and MLE utilize the full set of measured observables and therefore serve as benchmarks corresponding to full tomography.

\subsection{Structured Gibbs Reconstruction}
 GHZ states exhibit vanishing local expectation values and are primarily characterized by nonlocal correlations, and dominant global coherence captured by operators such as $X^{\otimes n}$~\cite{greenberger1989going, guhne2009entanglement}. Based on this structure, we define a hierarchy of models:

\begin{itemize}
    \item \textbf{G1 (local model):} Includes only single-qubit observables $\{X_i, Y_i, Z_i\}$. This model captures local properties but excludes all correlations.
    
    \item \textbf{G2 (nearest-neighbor model):} Extends G1 by including two-qubit nearest-neighbor correlations $\{X_i X_{i+1}, Y_i Y_{i+1}, Z_i Z_{i+1}\}$, capturing short-range structure.
    
    \item \textbf{G3 (global coherence model):} Further includes the global operators $X^{\otimes n}$ and $Y^{\otimes n}$, which directly probe the coherence between the $|0^n\rangle$ and $|1^n\rangle$ components of the GHZ state. These operators are consistent with the stabilizer structure of GHZ states and capture the dominant nonlocal coherence. While more general multi-qubit Pauli strings (e.g., mixed $X$--$Y$ strings) can also represent global correlations, they do not directly correspond to $|0^n\rangle$ and $|1^n\rangle$ components and therefore contribute only subdominantly. They are thus excluded at this level of the hierarchy.
    
    \item \textbf{G4 (extended correlation model):} Includes additional long-range two-qubit correlations beyond nearest neighbors, increasing expressivity without accessing the full operator space.
\end{itemize}

This hierarchy reflects a systematic inclusion of physically relevant correlations, progressing from local to global structure.

The number of parameters in each model is determined by the size of the operator set. For the systems studied in this work:

\begin{itemize}
    \item \textbf{3 qubits:} G1 = 9, G2 = 15, G3 = 17 (full: 63)
    \item \textbf{4 qubits:} G1 = 12, G2 = 21, G3 = 23, G4 = 35 (full: 255)
    \item \textbf{5 qubits:} G1 = 15, G2 = 27, G3 = 29, G4 = 50 (full: 1023)
\end{itemize}

The parameters $\{\lambda_k\}$ are obtained by minimizing the squared deviation between predicted and measured expectation values,
\begin{equation}
\mathcal{L} = \sum_k \left( \mathrm{Tr}(\rho P_k) - \langle P_k \rangle_{\mathrm{exp}} \right)^2,
\end{equation}
using the L-BFGS-B quasi-Newton optimization algorithm~\cite{byrd1995limited}.

\subsection{Evaluation Metrics}

The reconstructed states are evaluated using three complementary metrics.

The fidelity with respect to the target GHZ state is defined as~\cite{jozsa1994fidelity}
\begin{equation}
F(\rho, \rho_{\mathrm{GHZ}}) = \left( \mathrm{Tr} \sqrt{\sqrt{\rho_{\mathrm{GHZ}}} \rho \sqrt{\rho_{\mathrm{GHZ}}}} \right)^2.
\end{equation}

Agreement with MLE is quantified by computing the fidelity between each model and the MLE reconstruction.

The observable reconstruction error is defined as
\begin{equation}
\epsilon = \frac{1}{|\mathcal{P}|} \sum_{P \in \mathcal{P}} 
\left( \langle P \rangle_{\mathrm{MLE}} - \langle P \rangle_{\mathrm{model}} \right)^2.
\end{equation}

\subsection{Residual Analysis}

To identify correlations not captured by the SG--QST models, we compute residuals at the level of Pauli observables,
\begin{equation}
\Delta(P) = \langle P \rangle_{\mathrm{MLE}} - \langle P \rangle_{\mathrm{model}}.
\end{equation}

These residuals quantify the mismatch in observable space and are used to identify dominant correlations not included in the chosen operator set~\cite{torlai2018neural}.

Finally, the performance of all reconstruction methods is evaluated as a function of system size ($n = 3, 4, 5$), enabling direct comparison between full tomography and SG--QST models in terms of reconstruction accuracy and parameter efficiency.

%=========================================================

%=========================================================

%=========================================================
\section{Results}
\label{sec:Results}
\subsection{3-Qubit Reconstruction}

\begin{figure}[H]
    \centering
    % Top row - two figures
    \includegraphics[width=0.49\linewidth]{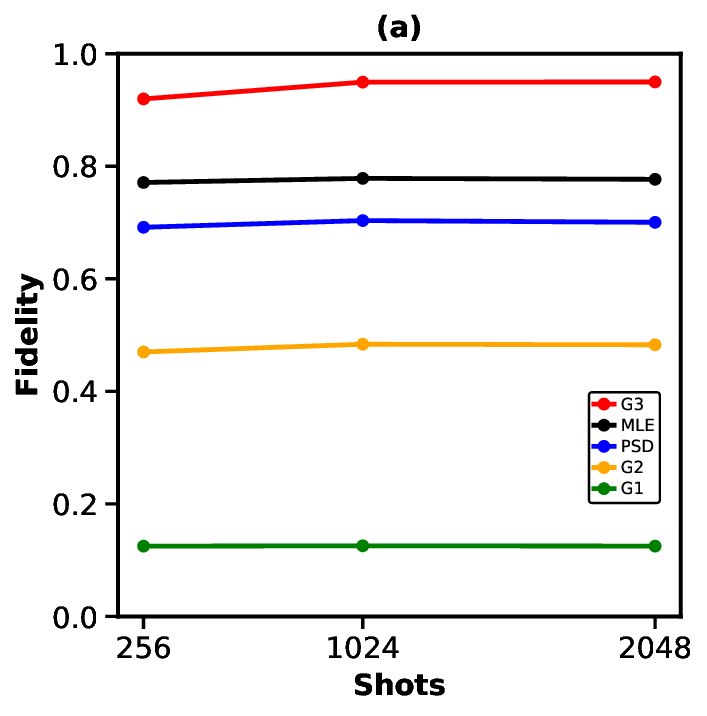}
    \hfill
    \includegraphics[width=0.49\linewidth]{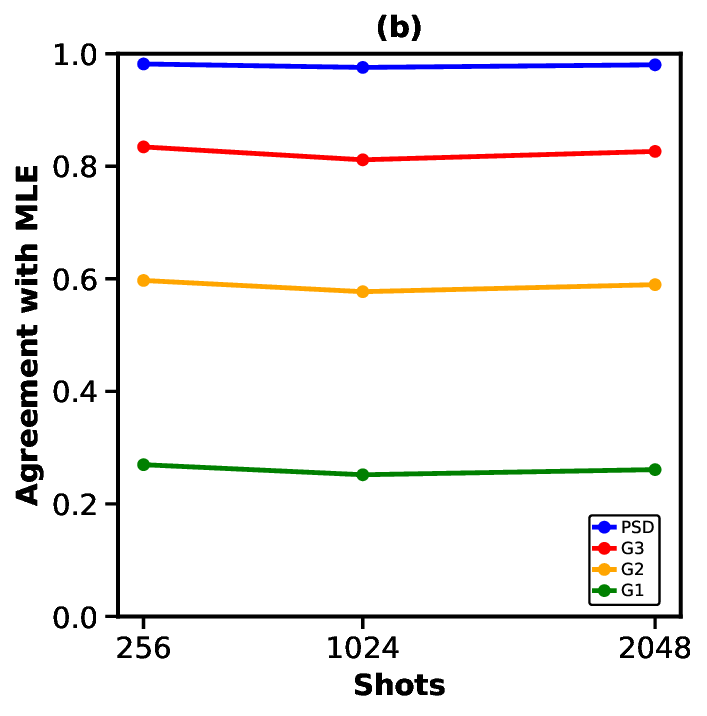}
    % Bottom row - one centered figure
    \includegraphics[width=0.49\linewidth]{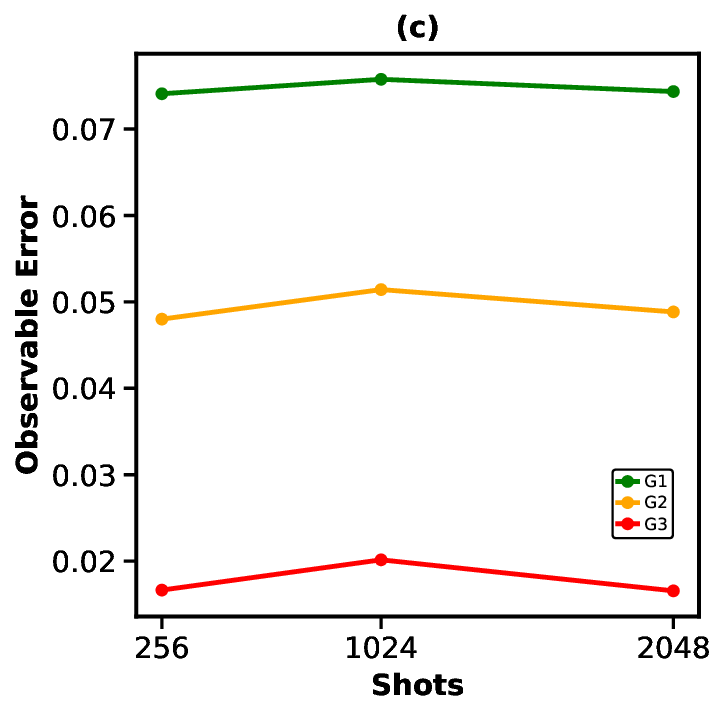}
    
    \caption{(a) Fidelity with respect to the ideal three-qubit GHZ state, (b) agreement with the MLE reconstruction, and (c) observable reconstruction error, shown as a function of measurement shots. The results demonstrate weak dependence on shot count and a clear hierarchy across models, with the G3 model achieving high accuracy using a reduced set of physically motivated observables.}
    \label{fig:3q_combined}
\end{figure}
\subsubsection{Fidelity with Target State}

The fidelity of the reconstructed density matrices with respect to the ideal three-qubit GHZ state is shown in Fig.~\ref{fig:3q_combined}(a). Across all shot values (256, 1024, and 2048), a clear hierarchy is observed among the models.

The G1 model (9 parameters), restricted to single-qubit observables, yields low fidelity (0.125, 0.126, 0.125), confirming that local information alone is insufficient to capture multipartite entanglement. Incorporating nearest-neighbor correlations, the G2 model (15 parameters) improves the fidelity to (0.470, 0.484, 0.483), indicating partial recovery of the state structure through short-range correlations.
\begin{table}[H]
\centering
\caption{Fidelity with respect to the ideal three-qubit GHZ state.}
\label{tab:3q_fidelity}
\begin{tabular}{c|ccccc}
\hline
Shots & MLE & PSD & G1 & G2 & G3 \\
\hline
256  & 0.771 & 0.692 & 0.125 & 0.470 & 0.920 \\
1024 & 0.778 & 0.703 & 0.126 & 0.484 & 0.949 \\
2048 & 0.777 & 0.700 & 0.125 & 0.483 & 0.950 \\
\hline
\end{tabular}
\end{table}
A significant improvement is observed for the G3 model (17 parameters), which includes global multi-qubit correlations. This model achieves high fidelity (0.920, 0.949, 0.950), with respect to the ideal state than MLE reconstruction (0.771, 0.778, 0.777), while using substantially fewer parameters than the full Pauli operator space ($4^3 - 1 = 63$).

In contrast, the PSD reconstruction yields fidelities (0.692, 0.703, 0.700), remaining consistently below the G3 model despite utilizing the full set of observables. This indicates that incorporating physically relevant global correlations is more effective than fitting all measured data indiscriminately.

Notably, the fidelity shows minimal variation with increasing shot count, indicating that reconstruction performance in this regime is primarily determined by model structure rather than statistical noise.

\subsubsection{Agreement with Maximum Likelihood Reconstruction}

The agreement of each model with the MLE reconstruction is shown in Fig.~\ref{fig:3q_combined}(b). As expected, the PSD reconstruction exhibits near-perfect agreement (0.982, 0.976, 0.980), reflecting its use of the full observable set.
\begin{table}[H]
\centering
\caption{Agreement with MLE reconstruction for the three-qubit system.}
\label{tab:3q_agreement}
\begin{tabular}{c|cccc}
\hline
Shots & PSD & G1 & G2 & G3 \\
\hline
256  & 0.982 & 0.270 & 0.597 & 0.834 \\
1024 & 0.976 & 0.252 & 0.577 & 0.811 \\
2048 & 0.980 & 0.261 & 0.590 & 0.826 \\
\hline
\end{tabular}
\end{table}
The structured models again show a clear progression. The G1 model yields low agreement (0.270, 0.252, 0.261), while the G2 model improves this to (0.597, 0.577, 0.590), capturing a significant portion of the experimentally reconstructed correlations.

The G3 model achieves substantially higher agreement (0.834, 0.811, 0.826), indicating that the inclusion of global multi-qubit correlations enables the structured model to reproduce the dominant features of the MLE state.

Importantly, while PSD maximizes agreement with MLE, it does not yield the highest fidelity with respect to the target GHZ state. In contrast, the G3 model achieves strong agreement while simultaneously maintaining high fidelity, indicating that it captures the physically relevant structure without overfitting to statistical fluctuations.

\subsubsection{Observable Reconstruction Error}

The observable reconstruction error with respect to the MLE estimate is shown in Fig.~\ref{fig:3q_combined}(c). A systematic reduction in error is observed with increasing model complexity.
\begin{table}[H]
\centering
\caption{Observable reconstruction error (with respect to MLE) for the three-qubit system.}
\label{tab:3q_error}
\begin{tabular}{c|ccc}
\hline
Shots & G1 & G2 & G3 \\
\hline
256  & 0.074 & 0.048 & 0.017 \\
1024 & 0.076 & 0.051 & 0.020 \\
2048 & 0.074 & 0.049 & 0.017 \\
\hline
\end{tabular}
\end{table}
The G1 model exhibits the largest error (0.074, 0.076, 0.074), confirming that single-qubit observables are insufficient. The G2 model reduces the error to (0.048, 0.051, 0.049), indicating that short-range correlations account for a significant portion of the observed structure.

The G3 model achieves a substantially lower error (0.017, 0.020, 0.017), demonstrating that global multi-qubit correlations are essential for accurately reproducing the observable structure of the state.

Notably, this improvement is achieved with only two additional parameters beyond G2, indicating that the gain is driven by the inclusion of physically relevant correlations rather than a simple increase in model size.

\subsubsection{Residual Analysis}

To further characterize the limitations of the structured model, we analyze the dominant residuals between the MLE estimate and the G3 reconstruction. We focus on the 2048-shot case, as similar behavior is observed across all shot values.
\begin{table}[H]
\centering
\caption{Top residual Pauli operators $\Delta(P)$ for three-qubit, G3 model (2048 shots).}
\label{tab:3q_residuals}
\begin{tabular}{c|ccccc}
\hline
Operator & IZY & IZX & XZI & YZI & XYY \\
\hline
Residual & 0.350 & 0.302 & 0.298 & 0.283 & 0.239 \\
\hline
\end{tabular}
\end{table}
The largest residuals are associated with mixed-axis Pauli operators, with IZY (0.350), IZX (0.302), XZI (0.298), YZI (0.283), and XYY (0.239) being the most prominent.

These terms are not included in the observable set defining the G3 model and therefore represent correlations beyond the structured ansatz. Despite this, the high fidelity (0.950) and low observable error (0.017) indicate that the dominant GHZ correlations are accurately captured.

The residuals therefore quantify secondary features of the experimental state, rather than indicating a failure of the model to reproduce its primary structure.

\subsection{4-Qubit Reconstruction}

\begin{figure}[H]
    \centering
    % Top row - two figures
    \includegraphics[width=0.49\linewidth]{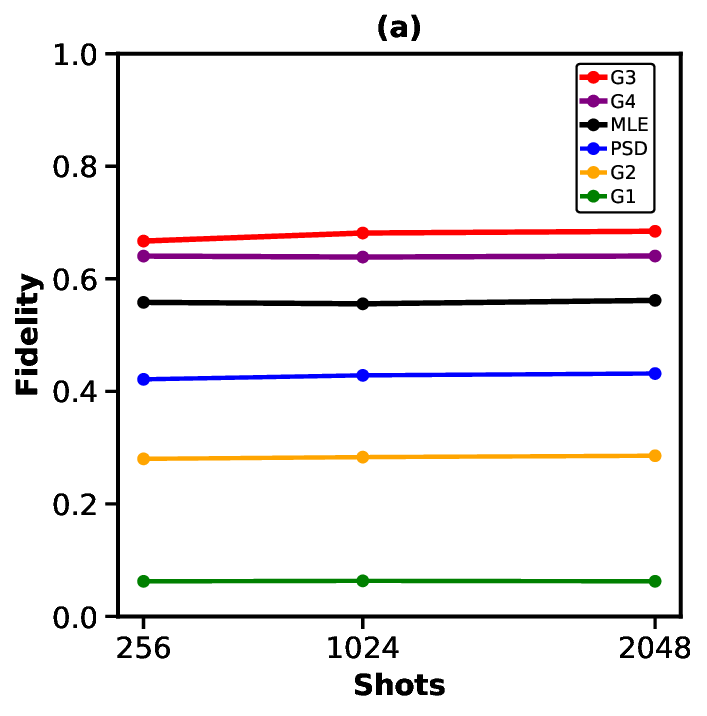}
    \hfill
    \includegraphics[width=0.49\linewidth]{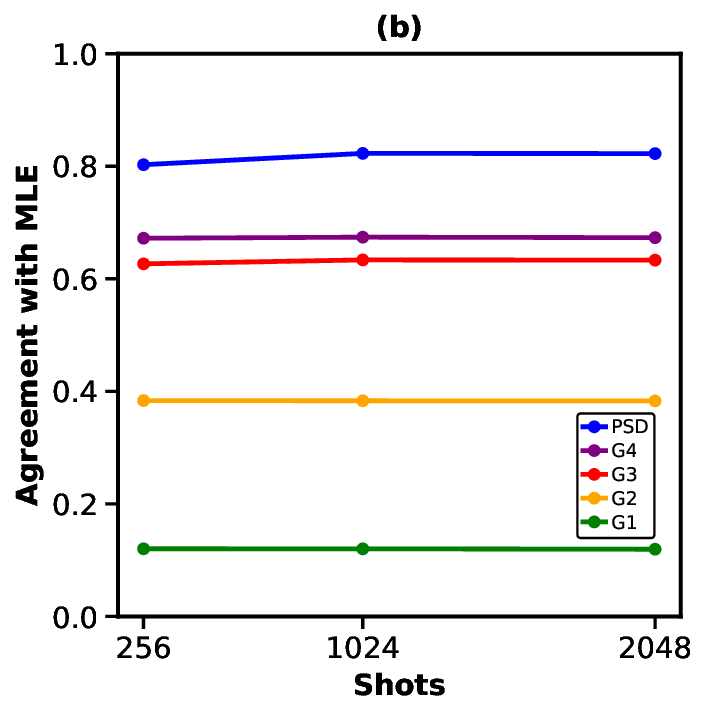}
    % Bottom row - one centered figure
    \includegraphics[width=0.49\linewidth]{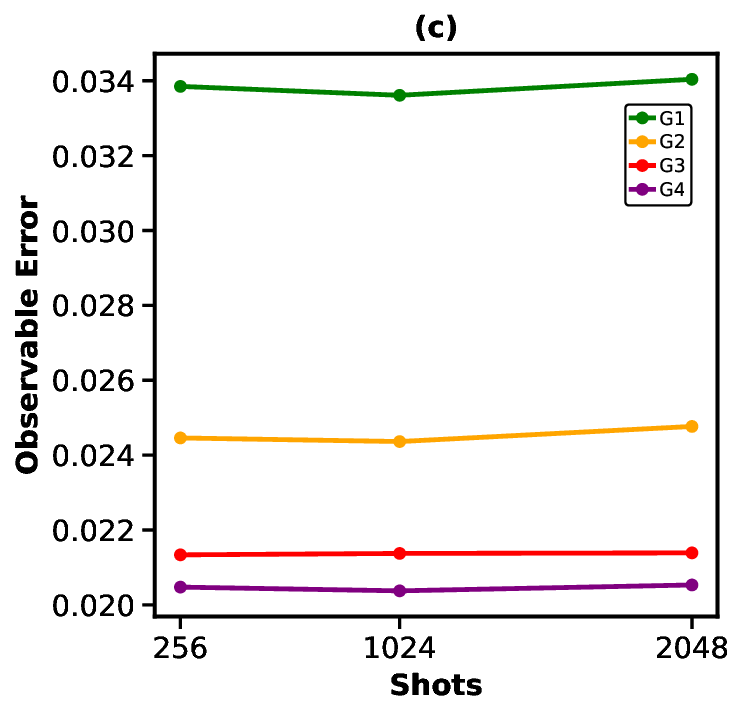}
    
    \caption{(a) Fidelity with respect to the ideal four-qubit GHZ state, (b) agreement with the MLE reconstruction, and (c) observable reconstruction error as a function of measurement shots.}
    \label{fig:4q_combined}
\end{figure}
\subsubsection{Fidelity with Target State}

We evaluate the fidelity of the reconstructed states with respect to the ideal four-qubit GHZ state for 256, 1024, and 2048 measurement shots. As shown in Fig.~\ref{fig:4q_combined}(a), a significant reduction in fidelity is observed compared to the three-qubit case, reflecting increased noise, circuit depth, and decoherence effects.

The MLE reconstruction, which effectively explores the full operator space (255 parameters for $n=4$), achieves fidelities of 0.558, 0.556, and 0.562 for 256, 1024, and 2048 shots, respectively. Despite its expressive power, the achieved fidelity remains limited by the quality of the experimental state.

The PSD reconstruction yields lower fidelities of 0.421, 0.428, and 0.432, indicating that consistency with measured observables alone does not guarantee recovery of the physically relevant state.

The structured models again exhibit a clear hierarchy. The G1 model (12 parameters) yields very low fidelity (0.062, 0.063, 0.062), confirming that local information alone is insufficient. The G2 model (21 parameters) improves the fidelity to 0.280, 0.283, and 0.286, capturing part of the short-range structure.
\begin{table}[H]
\centering
\caption{Fidelity with respect to the ideal four-qubit GHZ state.}
\label{tab:4q_fidelity}
\begin{tabular}{c|cccccc}
\hline
Shots & MLE & PSD & G1 & G2 & G3 & G4 \\
\hline
256  & 0.558 & 0.421 & 0.062 & 0.280 & 0.667 & 0.640 \\
1024 & 0.556 & 0.428 & 0.063 & 0.283 & 0.681 & 0.639 \\
2048 & 0.562 & 0.432 & 0.062 & 0.286 & 0.684 & 0.641 \\
\hline
\end{tabular}
\end{table}
The G3 model (23 parameters) achieves significantly higher fidelities of 0.667, 0.681, and 0.684, demonstrating that global coherence remains the dominant feature required to describe the GHZ state.

Extending the model to G4 (35 parameters) results in fidelities of 0.640, 0.639, and 0.641. The absence of a substantial improvement over G3 indicates that the dominant structure is already captured at the G3 level, and additional parameters primarily describe weaker correlations.

\subsubsection{Agreement with Maximum Likelihood Reconstruction}
To further evaluate the quality of the reconstructed states, we compute the fidelity between each model and the MLE reconstruction. As shown in Fig.~\ref{fig:4q_combined}(b), the PSD reconstruction shows high agreement with MLE, with values of 0.803, 0.823, and 0.825 for increasing shot counts.
\begin{table}[H]
\centering
\caption{Agreement with MLE reconstruction for the four-qubit system.}
\label{tab:4q_agreement}
\begin{tabular}{c|ccccc}
\hline
Shots & PSD & G1 & G2 & G3 & G4 \\
\hline
256  & 0.803 & 0.120 & 0.384 & 0.627 & 0.672 \\
1024 & 0.823 & 0.120 & 0.383 & 0.634 & 0.674 \\
2048 & 0.825 & 0.119 & 0.383 & 0.633 & 0.673 \\
\hline
\end{tabular}
\end{table}
The structured models exhibit a systematic improvement in agreement. The G1 model yields low agreement (0.120, 0.120, 0.119), while the G2 model improves this to 0.384, 0.383, and 0.383.

The G3 model achieves higher agreement values of 0.627, 0.634, and 0.633. Extending to G4 further improves the agreement to 0.672, 0.674, and 0.673.

However, the improvement from G3 to G4 remains small despite a substantial increase in parameters, indicating diminishing returns from including additional correlations.

\subsubsection{Observable Reconstruction Error}

We compute the observable reconstruction error with respect to the MLE estimate. As shown in Fig.~\ref{fig:4q_combined}(c), the error decreases systematically with model complexity.
\begin{table}[H]
\centering
\caption{Observable reconstruction error (with respect to MLE) for the four-qubit system.}
\label{tab:4q_error}
\begin{tabular}{c|cccc}
\hline
Shots & G1 & G2 & G3 & G4 \\
\hline
256  & 0.034 & 0.024 & 0.021 & 0.020 \\
1024 & 0.034 & 0.024 & 0.021 & 0.020 \\
2048 & 0.034 & 0.025 & 0.021 & 0.020 \\
\hline
\end{tabular}
\end{table}
The G1 model exhibits the largest error (0.034 across all shots), while the G2 model reduces the error to approximately 0.024–0.025.

The G3 model further reduces the error to approximately 0.021, indicating that global correlations are essential for accurately reproducing the observable structure.

Extending to G4 results in only a marginal reduction to 0.020, again showing that additional parameters provide limited improvement beyond G3.

\subsubsection{Residual Analysis}

We report the dominant residuals between the MLE estimate and the G3 model for 2048 measurement shots.

The largest residuals are observed for ZZZZ (-0.464), YYYY (-0.377), XXXX (-0.367), XXZZ (0.335), and YZII (0.333). These include both global multi-qubit terms and mixed-axis correlations that are not fully captured within the G3 observable set.
\begin{table}[H]
\centering
\caption{Top residual Pauli operators $\Delta(P)$ for the four-qubit,G3 model (2048 shots).}
\label{tab:4q_residuals}
\begin{tabular}{c|ccccc}
\hline
Operator & ZZZZ & YYYY & XXXX & XXZZ & YZII \\
\hline
Residual & -0.464 & -0.377 & -0.367 & 0.335 & 0.333 \\
\hline
\end{tabular}
\end{table}
Compared to the three-qubit case, the residual magnitudes are larger, consistent with the reduced fidelity (0.684) and higher observable error (0.021).

Despite these residuals, the G3 model captures the dominant structure of the GHZ state, while the residuals quantify additional correlations present in the experimental state.

\subsection{Five-Qubit Reconstruction}
\begin{figure}[H]
    \centering
    % Top row - two figures
    \includegraphics[width=0.49\linewidth]{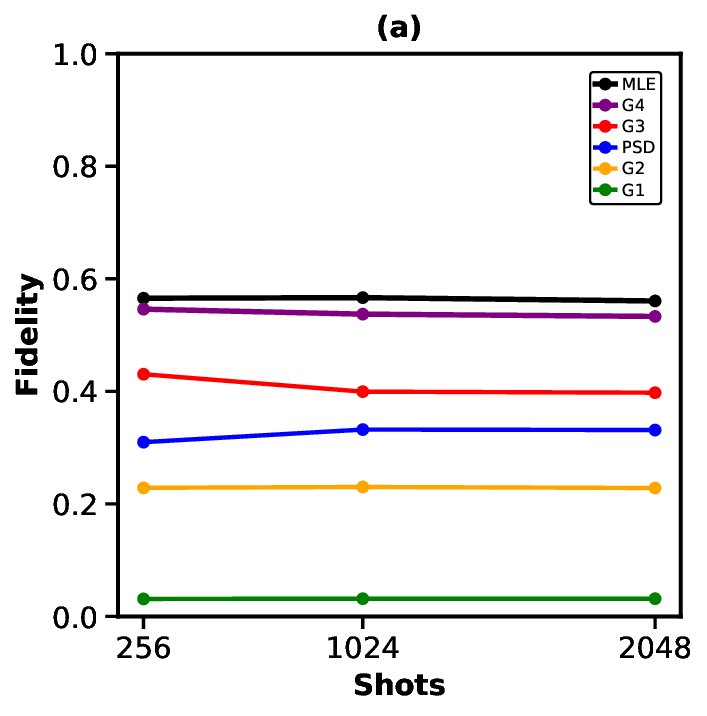}
    \hfill
    \includegraphics[width=0.49\linewidth]{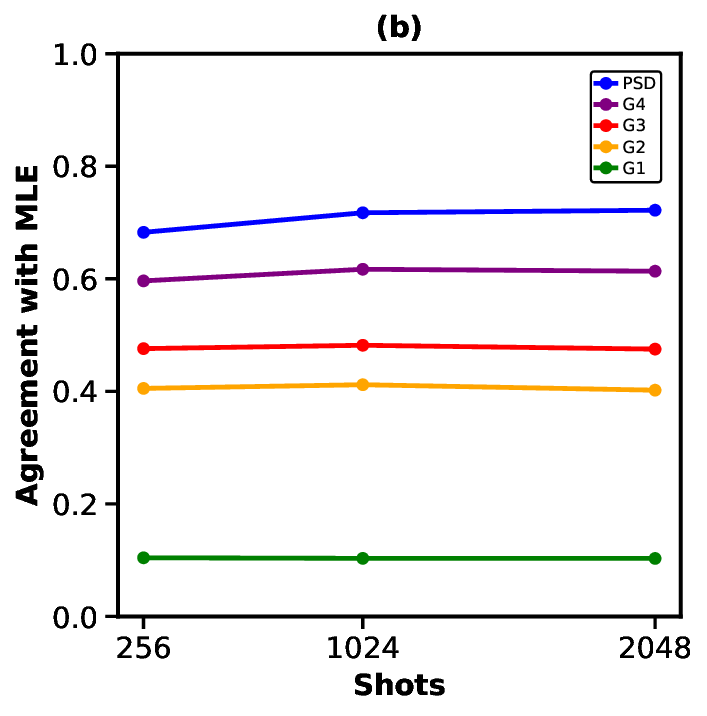}
    % Bottom row - one centered figure
    \includegraphics[width=0.49\linewidth]{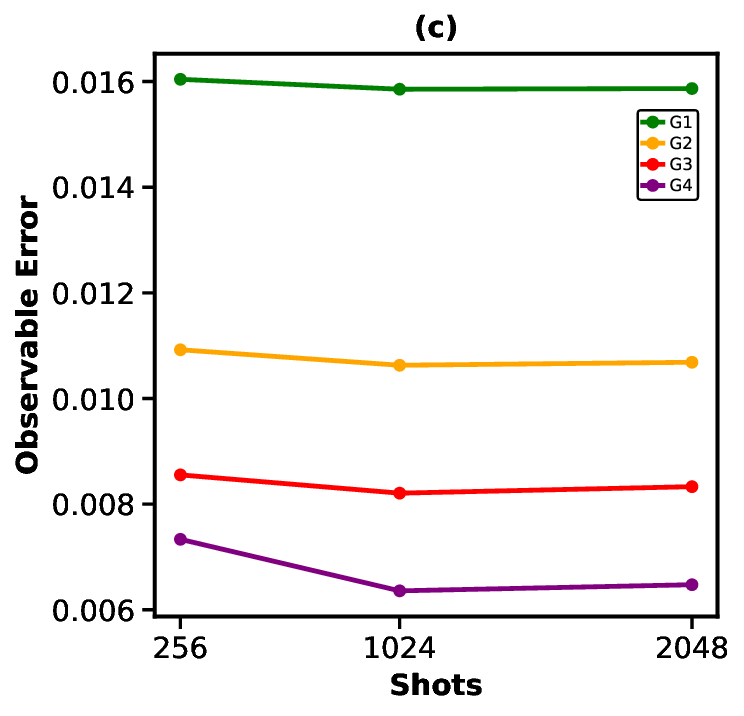}
    
    \caption{(a) Fidelity with respect to the ideal five-qubit GHZ state, (b) agreement with the MLE reconstruction, and (c) observable reconstruction error as a function of measurement shots.}
    \label{fig:5q_combined}
\end{figure}

\subsubsection{Fidelity with Target State}

We evaluate the fidelity of the reconstructed states with respect to the ideal five-qubit GHZ state for 256, 1024, and 2048 measurement shots. As shown in Fig.~\ref{fig:5q_combined}(a), the overall fidelity is reduced compared to smaller systems.

The MLE reconstruction yields fidelities of 0.573, 0.573, and 0.549, while the PSD reconstruction gives lower values of 0.310, 0.332, and 0.331.

The structured models exhibit a clear progression. The G1 model (15 parameters) yields low fidelity ($\sim 0.03$). The G2 model (27 parameters) improves the fidelity to $\sim 0.23$.

\begin{table}[H]
\centering
\caption{Fidelity with respect to the ideal five-qubit GHZ state.}
\label{tab:5q_fidelity}
\begin{tabular}{c|cccccc}
\hline
Shots & MLE & PSD & G1 & G2 & G3 & G4 \\
\hline
256  & 0.573 & 0.310 & 0.031 & 0.229 & 0.430 & 0.546 \\
1024 & 0.573 & 0.332 & 0.032 & 0.230 & 0.399 & 0.537 \\
2048 & 0.549 & 0.331 & 0.032 & 0.228 & 0.398 & 0.533 \\
\hline
\end{tabular}
\end{table}
The G3 model (29 parameters) achieves fidelities of 0.430, 0.399, and 0.398, which is insufficient to fully capture the dominant structure of the reconstructed state.

Further improvement is obtained with the G4 model (50 parameters), yielding fidelities of 0.546, 0.537, and 0.533.

This improvement arises from the inclusion of additional correlation structure rather than the full operator space ($4^n - 1 = 1023$ for $n=5$), indicating that a structured subset of observables captures a significant portion of the state even at larger system size.

\subsubsection{Agreement with Maximum Likelihood Reconstruction}

We evaluate the fidelity between each reconstructed state and the MLE estimate. As shown in Fig.~\ref{fig:5q_combined}(b), the PSD reconstruction shows strong agreement with MLE (0.675, 0.713, 0.747).
\begin{table}[H]
\centering
\caption{Agreement with MLE reconstruction for the five-qubit system.}
\label{tab:5q_agreement}
\begin{tabular}{c|ccccc}
\hline
Shots & PSD & G1 & G2 & G3 & G4 \\
\hline
256  & 0.675 & 0.098 & 0.385 & 0.465 & 0.586 \\
1024 & 0.713 & 0.105 & 0.400 & 0.471 & 0.599 \\
2048 & 0.747 & 0.131 & 0.422 & 0.476 & 0.598 \\
\hline
\end{tabular}
\end{table}

The structured models exhibit a systematic increase in agreement. The G1 model yields low agreement (0.098, 0.105, 0.131), while the G2 model improves this to (0.385, 0.400, 0.422). The G3 model further increases the agreement to (0.465, 0.471, 0.476).

The highest agreement is obtained for the G4 model (0.586, 0.599, 0.598), reflecting the inclusion of additional correlations.

However, this increase in agreement does not directly translate to improved fidelity with respect to the ideal GHZ state, indicating that agreement with MLE primarily reflects consistency with the reconstructed experimental state.

\subsubsection{Observable Reconstruction Error}

We evaluate the observable reconstruction error with respect to the MLE estimate. As shown in Fig.~\ref{fig:5q_combined}(c), the error decreases with increasing model complexity.
\begin{table}[H]
\centering
\caption{Observable reconstruction error (with respect to MLE) for the five-qubit system.}
\label{tab:5q_error}
\begin{tabular}{c|cccc}
\hline
Shots & G1 & G2 & G3 & G4 \\
\hline
256  & 0.016 & 0.011 & 0.009 & 0.007 \\
1024 & 0.016 & 0.011 & 0.009 & 0.007 \\
2048 & 0.016 & 0.011 & 0.009 & 0.007 \\
\hline
\end{tabular}
\end{table}
The G1 model yields the largest error (0.016), while the G2 model reduces it to approximately 0.011.

Further reduction is obtained for the G3 model (0.0087, 0.0087, 0.0086), and the lowest error is achieved by the G4 model (0.0074, 0.0068, 0.0069).

The reduction in error indicates improved agreement with the reconstructed state, although the improvement from G3 to G4 remains moderate compared to earlier transitions.

\subsubsection{Residual Analysis}

We examine the dominant residuals between the MLE reconstruction and the G3 model.
\begin{table}[H]
\centering
\caption{Top residual Pauli operators $\Delta(P)$ for the five-qubit, G3 model (2048 shots).}
\label{tab:5q_residuals}
\begin{tabular}{c|ccccc}
\hline
Operator & YXXYX & ZIIZI & ZIIIZ & IZIZI & YXXXY \\
\hline
$\Delta(P)$ & -0.151 & 0.146 & 0.139 & 0.103 & -0.102 \\
\hline
\end{tabular}
\end{table}
For the five-qubit system, the largest residuals are observed for mixed-axis Pauli operators such as YXXYX, ZIIZI, ZIIIZ, IZIZI, and YXXXY, with magnitudes up to $\sim 0.15$.

These operators are not included in the G3 model. Extending to G4 reduces some residuals, but heterogeneous Pauli strings remain among the dominant contributions.

Despite these residuals, the structured models reproduce the primary GHZ correlations, while the residuals quantify additional structure present in the experimental state.

\subsection{Scaling Behavior and Model Efficiency}\begin{figure}[H] \centering \includegraphics[width=0.90\linewidth]{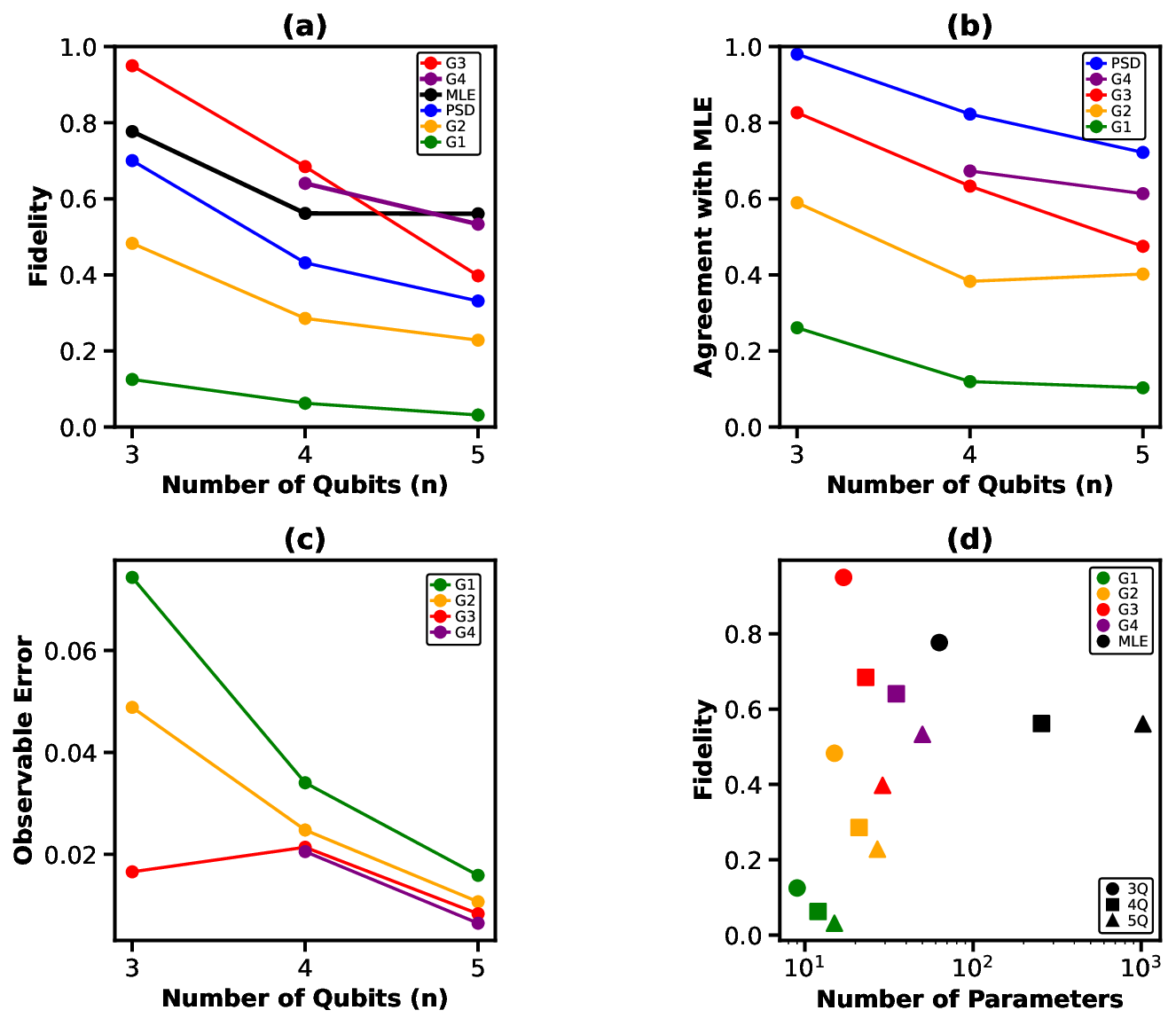} \caption{(a) Fidelity with respect to the ideal GHZ state, (b) agreement with the MLE reconstruction, (c) observable reconstruction error, and (d) fidelity as a function of the number of model parameters, shown for three-, four-, and five-qubit systems.} \label{fig:scaling} \end{figure}
\subsubsection{Fidelity Scaling}
As shown in Fig.~\ref{fig:scaling}(a), the fidelity with respect to the ideal GHZ state decreases with increasing system size for all reconstruction methods, reflecting increased noise, circuit depth, and decoherence.
The MLE reconstruction maintains relatively high fidelity across system sizes; however, this performance is obtained by optimizing over the full operator space, which scales exponentially ($4^n - 1$ parameters). In contrast, the structured models operate within a restricted parameter space.
Within the structured hierarchy, a clear pattern emerges. The transition from G2 to G3 yields a substantial improvement across all system sizes, driven by the inclusion of global coherence terms that capture the defining feature of GHZ states. In comparison, the transition from G3 to G4 produces only moderate gains, indicating that once the dominant correlations are included, additional parameters provide diminishing improvement.
This behavior demonstrates that reconstruction accuracy is not determined solely by the size of the parameter space, but by whether the selected observables capture the dominant physical structure of the state.
\subsubsection{Agreement with MLE}
Figure~\ref{fig:scaling}(b) shows the agreement with the MLE reconstruction. The PSD method exhibits the highest agreement, as it is derived from the same full set of measured observables.
The structured models show increasing agreement with model complexity, but decreasing agreement with system size, reflecting the increasing difficulty of reproducing the full reconstructed state within a restricted observable set.
Agreement with MLE should be interpreted alongside fidelity to the target state, as it reflects consistency with the reconstructed experimental density matrix rather than direct proximity to the ideal GHZ state.
\subsubsection{Observable Error Scaling}
The observable reconstruction error, shown in Fig.~\ref{fig:scaling}(c), decreases with increasing model complexity and increases with system size.
The G1 and G2 models exhibit large errors, confirming that local and short-range correlations are insufficient. The G3 and G4 models achieve significantly lower errors, indicating that global and higher-order correlations are essential for reproducing the dominant observable structure.
Notably, the reduction in error from G3 to G4 is relatively small compared to earlier transitions, further supporting that the dominant structure of the state is already captured at lower model complexity.
\subsubsection{Parameter Efficiency and Scaling Advantage}
Figure~\ref{fig:scaling}(d) captures the central result of this work by directly comparing reconstruction fidelity against the number of model parameters across system sizes.
While MLE achieves strong performance, it requires an exponentially scaling parameter space (63, 255, and 1023 parameters for three-, four-, and five-qubit systems, respectively). In contrast, the structured models achieve comparable performance using only a small fraction of these parameters.
For instance, in the five-qubit case, the G4 model approaches MLE-level fidelity while using only 50 parameters, representing an order-of-magnitude reduction in complexity. Importantly, this improvement is not simply a consequence of increasing parameter count. Instead, it arises from the targeted inclusion of physically relevant correlations.
The substantial performance gain observed when moving from G2 to G3 highlights the critical role of global coherence, while the comparatively smaller gain from G3 to G4 demonstrates diminishing returns once the dominant structure is captured.
These results establish that reconstruction accuracy is governed by the relevance of the chosen observables rather than the size of the parameter space. As system size increases, the required model complexity grows modestly, while remaining significantly below that of full tomography.
Overall, this demonstrates that structured reconstruction provides a scalable and physically grounded alternative to full quantum state tomography, enabling accurate descriptions of multipartite quantum states with substantially reduced parameter requirements.
\section{Conclusion}
\label{sec:Conclusion}
In this work, we introduced a structured approach to quantum state reconstruction based on a Gibbs parametrization with a systematically restricted set of observables. Rather than performing full quantum state tomography over an exponentially large operator space, the proposed method targets physically relevant correlations that define the underlying structure of the state.
Using GHZ states as a benchmark, we demonstrated that the dominant features of multipartite entanglement are captured by a small subset of observables, particularly those encoding global coherence. The structured models achieve a substantial fraction of the reconstruction accuracy obtained by full methods such as MLE, while requiring an order-of-magnitude fewer parameters.
The scaling analysis shows that reconstruction accuracy is governed primarily by the relevance of the selected observables rather than the size of the parameter space. In particular, the transition from local and short-range models to those incorporating global correlations yields the most significant improvement, while further increases in model complexity result in diminishing returns.
These results establish that accurate reconstruction of structured quantum states does not require access to the full operator space, but can be achieved through physically motivated model reduction. This provides a scalable alternative to conventional tomography for structured quantum states, with clear advantages in both computational efficiency and interpretability.
More broadly, the framework developed here suggests a strategy for quantum state reconstruction in larger systems: identify and prioritize the dominant correlation structure rather than attempting to reconstruct all degrees of freedom. This perspective is particularly relevant for near-term quantum devices, where noise, limited measurements, and system size constrain the applicability of full tomography.

\begin{acknowledgments}
The authors gratefully acknowledge the Central University
of Himachal Pradesh for providing the essential facilities and
support during the course of this research.
The authors thank Akshay Gaikwad  for their insightful discussions and suggestions. 
\end{acknowledgments}
%=========================================================
% Bibliography
%=========================================================

\bibliographystyle{apsrev4-2}
\bibliography{refrences}

@article{lvovsky2009continuous,
  title={Continuous-variable optical quantum-state tomography},
  author={Lvovsky, Alexander I and Raymer, Michael G},
  journal={Reviews of modern physics},
  volume={81},
  number={1},
  pages={299--332},
  year={2009},
  publisher={APS}
}

@article{rehacek2007diluted,
  title={Diluted maximum-likelihood algorithm for quantum tomography},
  author={{\v{R}}eh{\'a}{\v{c}}ek, Jaroslav and Hradil, Zden{\v{e}}k and Knill, E and Lvovsky, AI},
  journal={Physical Review A—Atomic, Molecular, and Optical Physics},
  volume={75},
  number={4},
  pages={042108},
  year={2007},
  publisher={APS}
}

@article{gross2010quantum,
  title={Quantum state tomography via compressed sensing},
  author={Gross, David and Liu, Yi-Kai and Flammia, Steven T and Becker, Stephen and Eisert, Jens},
  journal={Physical review letters},
  volume={105},
  number={15},
  pages={150401},
  year={2010},
  publisher={APS},
  doi=" 10.1103/PhysRevLett.105.150401"
}

@article{flammia2012quantum,
  title={Quantum tomography via compressed sensing: error bounds, sample complexity and efficient estimators},
  author={Flammia, Steven T and Gross, David and Liu, Yi-Kai and Eisert, Jens},
  journal={New Journal of Physics},
  volume={14},
  number={9},
  pages={095022},
  year={2012},
  publisher={IOP Publishing},
  doi="10.1088/1367-2630/14/9/095022"
}

@article{kalev2015quantum,
  title={Quantum tomography protocols with positivity are compressed sensing protocols},
  author={Kalev, Amir and Kosut, Robert L and Deutsch, Ivan H},
  journal={npj Quantum Information},
  volume={1},
  number={1},
  pages={15018},
  year={2015},
  publisher={Nature Publishing Group},
  doi="10.1038/npjqi.2015.18"
}

@article{cramer2010efficient,
  title={Efficient quantum state tomography},
  author={Cramer, Marcus and Plenio, Martin B and Flammia, Steven T and Somma, Rolando and Gross, David and Bartlett, Stephen D and Landon-Cardinal, Olivier and Poulin, David and Liu, Yi-Kai},
  journal={Nature communications},
  volume={1},
  number={1},
  pages={149},
  year={2010},
  publisher={Nature Publishing Group UK London},
  doi="10.1038/ncomms1147"
}

@article{lanyon2017efficient,
  title={Efficient tomography of a quantum many-body system},
  author={Lanyon, Ben P and Maier, Christiane and Holz{\"a}pfel, Milan and Baumgratz, Tillmann and Hempel, Cornelius and Jurcevic, Petar and Dhand, Ish and Buyskikh, AS and Daley, Andrew J and Cramer, Marcus and others},
  journal={Nature Physics},
  volume={13},
  number={12},
  pages={1158--1162},
  year={2017},
  publisher={Nature Publishing Group UK London},
  doi="10.1038/NPHYS4244"
}

@article{huszar2012adaptive,
  title={Adaptive Bayesian quantum tomography},
  author={Husz{\'a}r, Ferenc and Houlsby, Neil MT},
  journal={Physical Review A—Atomic, Molecular, and Optical Physics},
  volume={85},
  number={5},
  pages={052120},
  year={2012},
  publisher={APS},
  doi="https://doi.org/10.1103/PhysRevA.85.052120"
}

@article{carleo2017solving,
  title={Solving the quantum many-body problem with artificial neural networks},
  author={Carleo, Giuseppe and Troyer, Matthias},
  journal={Science},
  volume={355},
  number={6325},
  pages={602--606},
  year={2017},
  publisher={American Association for the Advancement of Science},
  doi="10.1126/science.aag2302"
}

@article{flammia2011direct,
  title={Direct fidelity estimation from few Pauli measurements},
  author={Flammia, Steven T and Liu, Yi-Kai},
  journal={Physical review letters},
  volume={106},
  number={23},
  pages={230501},
  year={2011},
  publisher={APS},
  doi="https://doi.org/10.1103/PhysRevLett.106.230501"
}

@article{da2011practical,
  title={Practical characterization of quantum devices without tomography},
  author={da Silva, Marcus P and Landon-Cardinal, Olivier and Poulin, David},
  journal={Physical Review Letters},
  volume={107},
  number={21},
  pages={210404},
  year={2011},
  publisher={APS},
  doi="https://doi.org/10.1103/PhysRevLett.107.210404"
}

@article{huang2020predicting,
  title={Predicting many properties of a quantum system from very few measurements},
  author={Huang, Hsin-Yuan and Kueng, Richard and Preskill, John},
  journal={Nature Physics},
  volume={16},
  number={10},
  pages={1050--1057},
  year={2020},
  publisher={Nature Publishing Group UK London},
  doi="https://doi.org/10.1038/s41567-020-0932-7"
}

@article{huang2021efficient,
  title={Efficient estimation of Pauli observables by derandomization},
  author={Huang, Hsin-Yuan and Kueng, Richard and Preskill, John},
  journal={Physical review letters},
  volume={127},
  number={3},
  pages={030503},
  year={2021},
  publisher={APS},
  doi="https://doi.org/10.1103/PhysRevLett.127.030503"
}

@inproceedings{aaronson2018shadow,
  title={Shadow tomography of quantum states},
  author={Aaronson, Scott},
  booktitle={Proceedings of the 50th annual ACM SIGACT symposium on theory of computing},
  pages={325--338},
  year={2018},
  doi=". https://doi.org/10.1145/
3188745.3188802"
}

@article{dur2000three,
  title={Three qubits can be entangled in two inequivalent ways},
  author={D{\"u}r, Wolfgang and Vidal, Guifre and Cirac, J Ignacio},
  journal={Physical Review A},
  volume={62},
  number={6},
  pages={062314},
  year={2000},
  publisher={APS}
}

@book{nielsen2002quantum,
  title={Quantum Computation and Quantum Information},
  author={Nielsen, Michael A. and Chuang, Isaac L.},
  year={2002},
  publisher={Cambridge University Press}
}

@book{paris2004quantum,
  title={Quantum State Estimation},
  author={Paris, Matteo G. A. and Rehacek, Jaroslav},
  year={2004},
  publisher={Springer}
}

@article{eisert2020quantum,
  title={Quantum certification and benchmarking},
  author={Eisert, Jens and Hangleiter, Dominik and Walk, Nathan and Roth, Ingo and Markham, Damian and Parekh, Rhea and Chabaud, Ulysse and Kashefi, Elham},
  journal={Nature Reviews Physics},
  volume={2},
  number={7},
  pages={382--390},
  year={2020},
  publisher={Nature Publishing Group UK London},
  doi="https://doi.org/10.1038/
s42254-020-0186-4"
}

@article{teo2012incomplete,
  title={Incomplete quantum state estimation: A comprehensive study},
  author={Teo, Yong Siah and Stoklasa, Bohumil and Englert, Berthold-Georg and {\v{R}}eh{\'a}{\v{c}}ek, Jaroslav and Hradil, Zden{\v{e}}k},
  journal={Physical Review A—Atomic, Molecular, and Optical Physics},
  volume={85},
  number={4},
  pages={042317},
  year={2012},
  publisher={APS},
  doi="https://doi.org/10.1103/PhysRevA.85.042317"
}

@book{greenberger1989going,
author="Greenberger, Daniel M.
and Horne, Michael A.
and Zeilinger, Anton",
editor="Kafatos, Menas",
title="Going Beyond Bell's Theorem",
bookTitle="Bell's Theorem, Quantum Theory and Conceptions of the Universe",
year="1989",
publisher="Springer Netherlands",
address="Dordrecht",
pages="69--72",
isbn="978-94-017-0849-4",
doi="10.1007/978-94-017-0849-4_10"
}

@article{guhne2009entanglement,
  title={Entanglement detection},
  author={G{\"u}hne, Otfried and T{\'o}th, G{\'e}za},
  journal={Physics Reports},
  volume={474},
  number={1-6},
  pages={1--75},
  year={2009},
  publisher={Elsevier},
  doi="https://doi.org/10.1016/j.physrep.2009.02.004"
}

@article{james2001measurement,
  title={Measurement of qubits},
  author={James, Daniel FV and Kwiat, Paul G and Munro, William J and White, Andrew G},
  journal={Physical Review A},
  volume={64},
  number={5},
  pages={052312},
  year={2001},
  publisher={APS},
  doi="https://doi.org/10.1103/PhysRevA.64.052312"
  
}

@article{hradil1997quantum,
  title={Quantum-state estimation},
  author={Hradil, Zdenek},
  journal={Physical Review A},
  volume={55},
  number={3},
  pages={R1561},
  year={1997},
  publisher={APS},
  doi="https://doi.org/10.1103/PhysRevA.55.R1561"
}

@article{jaynes1957information,
  title={Information theory and statistical mechanics},
  author={Jaynes, Edwin T},
  journal={Physical review},
  volume={106},
  number={4},
  pages={620},
  year={1957},
  publisher={APS},
  doi="https://doi.org/10.1103/PhysRev.106.620"
}

@book{petz2008quantum,
  title={Quantum Information Theory and Quantum Statistics},
  author={Petz, D{\'e}nes},
  year={2008},
  publisher={Springer}
}

@article{byrd1995limited,
  title={A limited memory algorithm for bound constrained optimization},
  author={Byrd, Richard H and Lu, Peihuang and Nocedal, Jorge and Zhu, Ciyou},
  journal={SIAM Journal on scientific computing},
  volume={16},
  number={5},
  pages={1190--1208},
  year={1995},
  publisher={SIAM}
}

@article{haah2017sample,
  title={Sample-optimal tomography of quantum states},
  author={Haah, Jeongwan et al.},
  journal={IEEE Transactions on Information Theory},
  volume={63},
  number={9},
  pages={5628--5641},
  year={2017},
  doi="https://doi.org/10.1145/2897518.2897585"
}

@article{jozsa1994fidelity,
  title={Fidelity for mixed quantum states},
  author={Jozsa, Richard},
  journal={Journal of modern optics},
  volume={41},
  number={12},
  pages={2315--2323},
  year={1994},
  publisher={Taylor \& Francis},
  doi="https://doi.org/10.1080/09500349414552171"
}

@article{torlai2018neural,
  title={Neural-network quantum state tomography},
  author={Torlai, Giacomo and Mazzola, Guglielmo and Carrasquilla, Juan and Troyer, Matthias and Melko, Roger and Carleo, Giuseppe},
  journal={Nature physics},
  volume={14},
  number={5},
  pages={447--450},
  year={2018},
  publisher={Nature Publishing Group UK London},
  doi="https://doi.org/10.1038/s41567-018-0048-5"
}

@article{pan2000,
  title={Experimental test of quantum nonlocality in three-photon Greenberger--Horne--Zeilinger entanglement},
  author={Pan, Jian-Wei and Bouwmeester, Dik and Daniell, Matthew and Weinfurter, Harald and Zeilinger, Anton},
  journal={Nature},
  volume={403},
  number={6769},
  pages={515--519},
  year={2000},
  publisher={Nature Publishing Group UK London}
}

@article{monz2011,
  title={14-qubit entanglement: Creation and coherence},
  author={Monz, Thomas and Schindler, Philipp and Barreiro, Julio T and Chwalla, Michael and Nigg, Daniel and Coish, William A and Harlander, Maximilian and H{\"a}nsel, Wolfgang and Hennrich, Markus and Blatt, Rainer},
  journal={Physical Review Letters},
  volume={106},
  number={13},
  pages={130506},
  year={2011},
  publisher={APS},
  doi="https://doi.org/10.1103/PhysRevLett.106.130506"
}

@article{pan2012,
  title={Multiphoton entanglement and interferometry},
  author={Pan, Jian-Wei and Chen, Zeng-Bing and Lu, Chao-Yang and Weinfurter, Harald and Zeilinger, Anton and {\.Z}ukowski, Marek},
  journal={Reviews of Modern Physics},
  volume={84},
  number={2},
  pages={777--838},
  year={2012},
  publisher={APS},
  doi="https://doi.org/10.1103/RevModPhys.84.777"
}

@article{hillery1999,
  title={Quantum secret sharing},
  author={Hillery, Mark and Bu{\v{z}}ek, Vladim{\'\i}r and Berthiaume, Andr{\'e}},
  journal={Physical Review A},
  volume={59},
  number={3},
  pages={1829},
  year={1999},
  publisher={APS},
  doi="https://doi.org/10.1103/PhysRevA.59.1829"
}

@article{song2017,
  title={10-qubit entanglement and parallel logic operations with a superconducting circuit},
  author={Song, Chao and Xu, Kai and Liu, Wuxin and Yang, Chui-ping and Zheng, Shi-Biao and Deng, Hui and Xie, Qiwei and Huang, Keqiang and Guo, Qiujiang and Zhang, Libo and others},
  journal={Physical review letters},
  volume={119},
  number={18},
  pages={180511},
  year={2017},
  publisher={APS},
  doi="https://doi.org/10.1103/PhysRevLett.119.180511"
}

@article{patel-prr-2026,
  title = {Selective and efficient quantum state tomography for multiqubit systems},
  author = {Patel, Aniket and Gaikwad, Akshay and Huang, Tangyou and Kockum, Anton Frisk and Abad, Tahereh},
  journal = {Phys. Rev. Res.},
  volume = {8},
  issue = {1},
  pages = {013339},
  numpages = {15},
  year = {2026},
  month = {Mar},
  publisher = {American Physical Society},
  doi = {10.1103/hynl-kxl2},
  url = {https://link.aps.org/doi/10.1103/hynl-kxl2}
}

@article{Gaikwad-qst-2025,
doi = {10.1088/2058-9565/ae0baa},
url = {https://doi.org/10.1088/2058-9565/ae0baa},
year = {2025},
month = {oct},
publisher = {IOP Publishing},
volume = {10},
number = {4},
pages = {045055},
author = {Gaikwad, Akshay and Torres, Manuel Sebastian and Ahmed, Shahnawaz and Kockum, Anton Frisk},
title = {Gradient-descent methods for fast quantum state tomography},
journal = {Quantum Science and Technology}
}

@Article{Gaikwad-qip-2022,
author={Gaikwad, Akshay
and {Arvind}
and Dorai, Kavita},
title={Efficient experimental characterization of quantum processes via compressed sensing on an NMR quantum processor},
journal={Quantum Information Processing},
year={2022},
month={Nov},
day={19},
volume={21},
number={12},
pages={388},
issn={1573-1332},
doi={10.1007/s11128-022-03695-3},
url={https://doi.org/10.1007/s11128-022-03695-3}
}

@article{gaikwad-pra-2024,
  title = {Neural-network-assisted quantum state and process tomography using limited data sets},
  author = {Gaikwad, Akshay and Bihani, Omkar and Arvind and Dorai, Kavita},
  journal = {Phys. Rev. A},
  volume = {109},
  issue = {1},
  pages = {012402},
  numpages = {9},
  year = {2024},
  month = {Jan},
  publisher = {American Physical Society},
  doi = {10.1103/PhysRevA.109.012402},
  url = {https://link.aps.org/doi/10.1103/PhysRevA.109.012402}
}

@Article{Gaikwad-qip-2021,
author={Gaikwad, Akshay
and {Arvind}
and Dorai, Kavita},
title={True experimental reconstruction of quantum states and processes via convex optimization},
journal={Quantum Information Processing},
year={2021},
month={Jan},
day={07},
volume={20},
number={1},
pages={19},
issn={1573-1332},
doi={10.1007/s11128-020-02930-z},
url={https://doi.org/10.1007/s11128-020-02930-z}
}

@article{gaikwad-epjd-2023,
author = {{A. Gaikwad} and {G. Singh} and {K. Dorai} and {Arvind}},
title = {{Direct tomography of quantum states and processes via weak measurements of Pauli spin operators on an NMR quantum processor}},
DOI= {10.1140/epjd/s10053-023-00791-6},
journal = {The European Physical Journal D},
year = {2023},
volume = {77},
pages = {209},
}

\end{document}